\documentclass[letterpaper, 10 pt, conference]{ieeeconf}  % Comment this line out if you need a4paper

\IEEEoverridecommandlockouts                              % This command is only needed if you want to use the \thanks command
\usepackage{graphicx}
\usepackage{url} % needed to auto break long url into lines with natbib
\usepackage{amsmath,amssymb,amsfonts}
\usepackage{algorithmic}
\usepackage{textcomp}
\usepackage{xcolor}
\usepackage[short]{optidef}
\usepackage{tikz}
\usetikzlibrary{arrows,arrows.meta,shapes,positioning,shadows,trees}
\usepackage{siunitx}
\usepackage{lipsum}
\usepackage{systeme}
\usepackage{siunitx}
\newcommand*\titleheader[1]{\gdef\@titleheader{#1}}

\NewDocumentCommand{\vartwo}{ >{\SplitArgument{2}{,}}m }{ \finalvartwo#1 }
\NewDocumentCommand{\finalvartwo}{mmm}{ \ensuremath{#1_\text{#2}} }
\NewDocumentCommand{\varthree}{ >{\SplitArgument{3}{,}}m }{ \finalvarthree#1 }
\NewDocumentCommand{\finalvarthree}{mmmm}{ \ensuremath{#1_{\text{#2},#3}} }

\title{\LARGE \bf
Convex Optimization for Fuel Cell Hybrid Trains: \\ Speed, Energy Management System, and Battery Thermals
}

\author{Rabee Jibrin$^*$, Stuart Hillmansen, Clive Roberts% <-this % stops a space
\thanks{All authors are with the Department of Electronic, Electrical and Systems Engineering and the Birmingham Centre for Railway Research and Education, University of Birmingham, United Kingdom}
\thanks{$^*$Corresponding author: {\tt\small rxj956@bham.ac.uk}}
}

\begin{document}

\onecolumn
{\color{blue} This paper has been submitted to the European Control Conference 2022, London, UK}
\twocolumn
\newpage

\maketitle
\thispagestyle{empty}
\pagestyle{empty}

%%%%%%%%%%%%%%%%%%%%%%%%%%%%%%%%%%%%%%%%%%%%%%%%%%%%%%%%%%%%%%%%%%%%%%%%%%%%%%%%
\begin{abstract}
We optimize the operation of a fuel cell hybrid train using convex optimization. The main objective is to minimize hydrogen fuel consumption for a target journey time while considering battery thermal constraints. The state trajectories: train speed, energy management system, and battery temperature, are all optimized concurrently within a single optimization problem. A novel thermal model is proposed in order to include battery temperature yet maintain formulation convexity. Simulations show fuel savings and better thermal management when temperature is optimized concurrently with the other states rather than sequentially---separately afterwards. The fuel reduction is caused by reduced cooling effort which is motivated by the formulation's awareness of active cooling energy consumption. The benefit is more pronounced for warmer ambient temperatures that require more cooling.
\end{abstract}
%%%%%%%%%%%%%%%%%%%%%%%%%%%%%%%%%%%%%%%%%%%%%%%%%%%%%%%%%%%%%%%%%%%%%%%%%%%%%%%%

\section{Introduction}

\subsection{Motivation}

\begin{figure}[!t]
\centering
  \begin{tikzpicture}
  \draw[black, very thick] (-0.25,0) rectangle (2,1) node[pos=0.5] {Battery};
  \draw[black, very thick] (-0.25,-1.5) rectangle (2,-0.5) node[pos=0.5] {PEMFC};
  \draw[black, very thick] (2.75,-1.5) rectangle (4.25,1) node[pos=0.5, align=center] {DC Link};
  \draw[black, very thick] (5.5,0.5) circle (0.5cm) node {MG};
  \draw[black, very thick] (5,-1.5) rectangle (6.5,-0.5) node[pos=0.5] {Auxiliary};
  \draw[>=triangle 45, <->] (2,0.5) -- (2.75,0.5); %between battery and dc link
  \draw[>=triangle 45, ->] (2,-1) -- (2.75,-1); %between fuel cell and dc link
  \draw[>=triangle 45, <->] (4.25,0.5) -- (5,0.5); %between dc link and MG
  \draw[>=triangle 45, ->] (4.25,-1) -- (5,-1); %between dc link and auxiliary
  \end{tikzpicture}
\caption{Fuel cell hybrid powertrain. Arrows depict feasible power flow.}
\label{fig:fc_powertrain}
\end{figure}
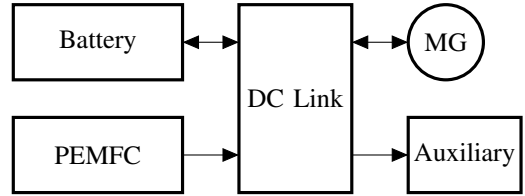

Hydrogen fuel cell hybrid trains (see Fig. \ref{fig:fc_powertrain}) are expected to play a key role in decarbonizing the railways owing to their lack of harmful emissions at point-of-use and adequate driving range; however, their total cost of ownership is projected to be higher than incumbent diesel trains primarily due to the higher cost of hydrogen fuel in comparison to diesel fuel \cite{RN200}. We aim at reducing hydrogen fuel consumption by optimizing train operation. Furthermore, battery thermals are included in order to extend battery lifetime. Convex optimization is used to alleviate computational concerns.

\subsection{Background}

Train speed optimization has been researched extensively owing to the large contribution of traction power towards rail energy consumption \cite{RN707}. More recently, the 2019 IEEE VTS Motor Vehicles Challenge brought attention to the energy management system (EMS) of fuel cell hybrid trains \cite{RN876}. The EMS determines power distribution among a hybrid vehicle's multiple power sources and is thus a vital determinant of efficiency. An extensive literature review of fuel cell hybrid EMS has been covered by \cite{RN318}. Simulations suggest that optimization-based algorithms outperform their rule-based counterparts which motivates our focus on the former \cite{RN565}.

While the aforementioned references are exclusively concerned with either train speed or EMS, other works have attempted to optimize both within a single optimization problem (concurrently) in order to achieve better solution optimality by embedding knowledge of the dynamic coupling between both trajectories, e.g., dynamic programming \cite{RN857}, indirect optimal control \cite{RN613}, integer programming \cite{RN924}, and relaxed convex optimization \cite{RN886,mine_vppc}.

The high capital cost of traction batteries has also motivated many to consider penalizing \cite{RN380} or bounding \cite{RN384} battery degradation, though strictly within the EMS problem setting---speed is optimized beforehand separately. The semi-empirical battery degradation model presented by \cite{RN370} as a function of temperature, state-of-charge, and C-rate, is the most often used. A common assumption among optimization formulations that consider battery degradation is an active cooling system that maintains a constant battery temperature which simplifies the degradation model to static temperature. This simplification can lead to unexpected battery degradation when subject to non-ideal thermal management in the real-world \cite{Filippi}. Therefore, dropping the static temperature assumption could further benefit battery lifetime, especially in light of experimental results that designate elevated temperatures as the leading cause of battery degradation \cite{RN386}. Moreover, including thermal constraints while planning the duty cycle can reduce the reliance on the active cooling system and thus reduce its parasitic energy draw \cite{RN926}. Formulations that did consider battery temperature as a bounded dynamic state have done so strictly within the EMS problem setting and often at great computational cost, e.g., genetic algorithm \cite{RN296}, dynamic programming \cite{RN669}, and relaxed convex optimization \cite{RN808}.

\subsection{Contribution and Outline}

The benefit from concurrent speed and EMS optimization has already been proven by previous works, yet literature lacks a concurrent formulation that considers battery thermal constraints. The high predictability of railway environments promises substantial returns for such a holistic optimization approach that considers many variables \textit{a priori}. We propose a relaxed convex optimization formulation that embeds a novel thermal model to achieve this goal. This paper builds upon a previous paper of ours that exclusively derived mathematical models and adds to it simulation results \cite{mine_mathmod}. To the the best of our knowledge, this is the first attempt at optimizing all mentioned state variables concurrently.

Section 2 presents the train model, section 3 details the problem formulation, section 4 analyzes simulation results.

\section{Modeling}

Common among model-based optimization for dynamic systems is to model the system in the time-domain, i.e., the model predicts system state after a temporal interval of $\vartwo{\Delta,t}$ seconds. However, a complication from optimizing vehicle speed in the time-domain is interpolating track information, e.g., gradient, when the physical location for a given temporal interval is dependent on the optimized speed and thus unknown \textit{a priori}. We address this by using the space-domain instead, i.e., the model predicts system state after a spatial interval of $\vartwo{\Delta,s}$ meters longitudinally along the track. As such, one can accurately retrieve track information for any interval by directly referring to its location in space. Herein, the models and thus the optimization problem are formulated in the discrete space-domain with zero-order hold between spatial intervals $\{\varthree{\Delta,s,i}|i=0,\cdots,N-1\}$.

Below, the train's longitudinal dynamics are first derived after which the powertrain components are covered. Figure \ref{fig:fc_powertrain} depicts the powertrain considered, a polymer electrolyte membrane fuel cell (PEMFC) in a hybrid configuration with a lithium-ion battery that power the motor-generator (MG) and train auxiliary loads. More details and figures of the models can be found in \cite{mine_mathmod}.

\subsection{Train Longitudinal Speed}

The train is assumed a point mass $m$ with an equivalent inertial mass $\vartwo{m,eq}$ traveling at a longitudinal speed $v$ that is controlled by traction motor force $\vartwo{F,m}$ and mechanical brakes force $\vartwo{F,brk}$ \cite{RN728}. The external forces acting on the train $\vartwo{F,ext}$ are represented by the summation of the Davis Equation $a+bv_i+cv_i^2$ and gravitational pull $mg\sin(\theta_i)$. To predict speed after a single spatial interval, construct
\begin{equation}\label{eq:kinetic_energy_1}
\frac{1}{2}\vartwo{m,eq}v_{i+1}^2 = \frac{1}{2}\vartwo{m,eq}v_{i}^2 + (\varthree{F,m,i} + \varthree{F,brk,i})\Delta_{\text{s},i} - \varthree{F,ext,i}\Delta_{\text{s},i}
\end{equation}
using kinetic energy $\vartwo{E,k.e.}=1/2 \vartwo{m,eq} v^2$, mechanical work $\vartwo{E,work}=F\Delta_\text{s}$, and the principle of energy conservation. Equation \eqref{eq:kinetic_energy_1} is nonlinear in $v$ but can be linearized by substituting the quadratic terms $v^2$ with $z$ and keeping the non-quadratic terms $v$ unchanged, namely
\begin{equation}\label{eq:kinetic_energy_2}
\frac{1}{2}\vartwo{m,eq}z_{i+1} = \frac{1}{2}\vartwo{m,eq}z_{i} + (\varthree{F,m,i} + \varthree{F,brk,i})\Delta_{\text{s},i} - \varthree{F,ext,i}\Delta_{\text{s},i}
\end{equation}
and
\begin{equation}\label{eq:f_ext}
\varthree{F,ext,i} = a+bv_i+cz_i + mg\sin(\theta_i). 
\end{equation}

The linear model \eqref{eq:kinetic_energy_2} relies on both $v$ and $z$ to define train speed and thus requires the non-convex equality constraint $v^2=z$ to hold true which is subsequently relaxed into the convex inequality
\begin{equation}\label{eq:relaxed_v_2}
v^2 \leq z.
\end{equation}

\subsection{Journey Time}

Total journey time is expressed as summation of time required for all spatial intervals $\sum_{i=0}^{N-1} \varthree{\Delta,s,i}/v_i$ but is non-linear in $v$. This expression can be replaced by the linear 
\begin{equation}
\sum_{i=0}^{N-1} \varthree{\Delta,s,i}\lambda_{v,i}
\end{equation} 
when used along the auxiliary non-convex equality $\lambda_v = 1/v$ which is then relaxed into the convex inequality
\begin{equation}\label{eq:lambda_v}
\lambda_{v} \geq 1/v
\end{equation}
for $v,\lambda_v > 0$ \cite{boyd2004convex}. Section 3 explains how the strict positivity constraint imposed on speed has a negligible impact on solution optimality and how the relaxed inequalities \eqref{eq:relaxed_v_2} and \eqref{eq:lambda_v} hold with equality at the optimal solution.

\subsection{Traction Motor}\label{sec:motor}

The electric power flow in Fig. \ref{fig:fc_powertrain} is described by 
\begin{equation}\label{eq:power_balance}
\vartwo{P,m}/\vartwo{\eta,m}(\vartwo{P,m}) + \vartwo{P,cool} + \vartwo{P,aux} = \vartwo{P,fc} + \vartwo{P,batt},
\end{equation}
where $\vartwo{P,m}$ is motor mechanical power, $\eta_m(\vartwo{P,m})$ is motor efficiency and thus $\vartwo{P,m}/\eta_m(\vartwo{P,m})$ is electric power at motor terminals, $\vartwo{P,cool}$ is the electric load of the battery cooling system, $\vartwo{P,aux}$ is other auxiliary loads, $\vartwo{P,fc}$ is fuel cell electric power output, and $\vartwo{P,batt}$ is battery electric power output. $\vartwo{P,aux}$ is modeled as constant.

The power balance expression \eqref{eq:power_balance} requires the non-convex constraint $\vartwo{P,m}=\vartwo{F,m}v$ to hold true in order to use it in conjunction with the speed model \eqref{eq:kinetic_energy_2}. To resolve this non-convexity, start by dividing \eqref{eq:power_balance} by $v$ to yield
\begin{equation}\label{eq:force_balance}
\vartwo{F,m}/\vartwo{\eta,m}(\vartwo{F,m},z) + \vartwo{F,cool} + \vartwo{P,aux}\lambda_v = \vartwo{F,fc} + \vartwo{F,batt},
\end{equation}
where motor efficiency is defined as $\vartwo{\eta,m}(\vartwo{F,m},z)$ instead of $\vartwo{\eta,m}(\vartwo{P,m})$, recall $P=F\sqrt{z}$. The alternative model \eqref{eq:force_balance} expresses energy flow per longitudinal meter traveled, recall $\vartwo{E,work}=F\vartwo{\Delta,s}$ and $F=P\lambda_v$. The forces $\vartwo{F,fc}$ and $\vartwo{F,batt}$ are fictitious but numerically represent the energy contribution of each power source per meter. Since \eqref{eq:force_balance} is directly written in terms of $\vartwo{F,m}$ the non-convex constraint $\vartwo{P,m}=\vartwo{F,m}v$ is dropped.

The equality \eqref{eq:force_balance} is non-convex due to the non-linearity in $\vartwo{F,m}/\vartwo{\eta,m}(\vartwo{F,m},z)$. Moreover, motor efficiency, $\vartwo{\eta,m}$, is typically a discrete look-up table rather than a smooth function. $\vartwo{F,m}/\vartwo{\eta,m}(\vartwo{F,m},z)$ can be accurately approximated by the convex quadratic polynomial $\vartwo{q,m}(\vartwo{F,m},z):=p_{00}+p_{10}z+p_{01}\vartwo{F,m}+p_{11}\vartwo{F,m}v+p_{20}z^2+p_{02}\vartwo{F,m}^2$, as shown by \cite[Fig. 2]{mine_mathmod}, which can be used to relax \eqref{eq:force_balance} into the convex inequality
\begin{equation}\label{eq:force_balance_polynomial}
\vartwo{q,m}(\vartwo{F,m},z) + \vartwo{F,cool} + \vartwo{P,aux}\lambda_v \leq \vartwo{F,fc} + \vartwo{F,batt}.
\end{equation}

Motors are constrained by two regions, a constant force region under the cutoff speed
\begin{equation}
\underline{\vartwo{F,m}} \leq\ \vartwo{F,m} \leq \overline{\vartwo{F,m}}
\end{equation}
and a constant power region above the cutoff speed
\begin{subequations}
\begin{alignat}{1}
\underline{\vartwo{P,m}}\lambda_v & \leq \vartwo{F,m},\\ 
\vartwo{F,m} & \leq \overline{\vartwo{P,m}}\lambda_v.
\end{alignat}
\end{subequations}

\subsection{Fuel Cell}\label{sec:fuel_cell}

To penalize hydrogen fuel consumption, we derive an expression for fuel energy consumed per longitudinal meter traveled. The exact fuel penalty per meter is $\vartwo{F,fc}/\vartwo{\eta,fc}(\vartwo{F,fc},z)$ which can be accurately approximated by the convex quadratic polynomial \cite[Fig. 3]{mine_mathmod}
\begin{equation}\label{eq:q_fc}
\begin{split}
\vartwo{q,fc}(\vartwo{F,fc},z) := &p_{00}'+p_{10}'z+p_{01}'\vartwo{F,fc}\\
&+p_{11}'\vartwo{F,fc}v+p_{20}'z^2+p_{02}'\vartwo{F,fc}^2.
\end{split}
\end{equation}

The fuel cell power constraints are
\begin{subequations}
\begin{alignat}{1}
\underline{\vartwo{P,fc}}\lambda_v & \leq \vartwo{F,fc},\\ 
\vartwo{F,fc} & \leq \overline{\vartwo{P,fc}}\lambda_v,
\end{alignat}
\end{subequations}
where $\underline{\vartwo{P,fc}}$ may be strictly positive to curtail the excessive degradation that accompanies idling and restarting \cite{RN170}.

\subsection{Battery State-of-Charge}\label{sec:battery_soc}

Predicting the battery's state-of-charge $\zeta$ is vital in order to guarantee charge-sustaining operation---terminal battery charge identical to initial. The battery is modeled with a fixed open-circuit voltage $\vartwo{U,oc}$ and a fixed internal resistance $R$, a model that is accurate for the narrow state-of-charge range employed by hybrid vehicles and is validated by \cite{RN700}.

The change of state-of-charge is
\begin{equation}\label{eq:dzeta_dt}
\Delta_\zeta = \frac{\vartwo{U,oc}-\sqrt{\vartwo{U,oc}^2-4\vartwo{P,batt}R}}{2R} \cdot \frac{1}{3600Q} \cdot \Delta_\text{t},
\end{equation}
where $Q$ is battery charge capacity, valid for $\vartwo{P,batt} \leq \vartwo{U,oc}^2/4R$ \cite{RN736}. Accordingly, a positive/(negative) $\vartwo{P,batt}$ will discharge/(charge) the battery
\begin{equation}\label{eq:zeta}
\zeta_{i+1}=\zeta_i-\Delta_{\zeta,i}.
\end{equation}

For a given $\vartwo{\Delta,t}$, \eqref{eq:dzeta_dt} is convex in $\vartwo{P,batt}$ which empowers the convex quadratic $q_\zeta(\vartwo{P,batt}):=\alpha \vartwo{P,batt}^2 + \beta \vartwo{P,batt}$ to accurately approximate it, as shown by \cite[Fig. 4]{mine_mathmod}. Nevertheless, an expression written in terms of spatial intervals $\vartwo{\Delta,s}$ rather than temporal intervals $\vartwo{\Delta,t}$ needs to be derived for a space-domain formulation. Start by assuming $\Delta_\zeta =q_\zeta(\vartwo{P,batt})\vartwo{\Delta,t}$
\begin{equation}
\Delta_\zeta = (\alpha\vartwo{P,batt}^2 + \beta \vartwo{P,batt})\vartwo{\Delta,t}
\end{equation}
which can be rewritten in terms of $\vartwo{P,batt}=\vartwo{F,batt}v$
\begin{equation}\label{eq:F_batt_original}
\Delta_\zeta = (\alpha\vartwo{F,batt}^2v^2 + \beta \vartwo{F,batt}v)\vartwo{\Delta,t}
\end{equation}
followed by the substitution $v=\Delta_\text{s}/\Delta_\text{t}$
\begin{equation}
\Delta_\zeta = \Big(\alpha\vartwo{F,batt}^2v \frac{\Delta_\text{s}}{\Delta_\text{t}} + \beta \vartwo{F,batt} \frac{\Delta_\text{s}}{\Delta_\text{t}}\Big)\vartwo{\Delta,t}
\end{equation}
then cancel out $\Delta_\text{t}$ in order to obtain the spatial expression
\begin{equation}\label{eq:battery_model_1}
\Delta_\zeta = \alpha\vartwo{F,batt}^2v\vartwo{\Delta,s} + \beta \vartwo{F,batt}\vartwo{\Delta,s}.
\end{equation}

Equation \eqref{eq:battery_model_1} is non-convex but can be rewritten as
\begin{equation}\label{eq:battery_model_2}
\alpha\vartwo{F,batt}^2\vartwo{\Delta,s} = \frac{\Delta_\zeta-\beta \vartwo{F,batt}\vartwo{\Delta,s}}{v}
\end{equation}
then subsumed into
\begin{equation}\label{eq:battery_model_3}
\alpha\vartwo{F,batt}^2\vartwo{\Delta,s} = \lambda_\zeta\lambda_v
\end{equation}
using the linear auxiliary constraint
\begin{equation}\label{eq:lambda_zeta}
\lambda_\zeta=\Delta_\zeta-\beta \vartwo{F,batt}\vartwo{\Delta,s}
\end{equation}
and the convex constraint \eqref{eq:lambda_v}. 

The relaxation of the non-convex equality \eqref{eq:battery_model_3},
\begin{equation}\label{eq:battery_model_4}
\alpha\vartwo{F,batt}^2\vartwo{\Delta,s} \leq \lambda_\zeta\lambda_v,
\end{equation}
forms a convex feasible set for $\lambda_\zeta,\lambda_v \geq 0$ which is nonrestrictive, since $\lambda_v$ and the left-hand side of \eqref{eq:battery_model_4} are non-negative by definition. Section 3 proves that the inequality \eqref{eq:battery_model_4} holds with equality at the optimal solution.

\subsection{Battery Temperature}\label{sec:battery_temp}

Battery temperature $\vartwo{T,batt}$ is to be modeled in order to keep temperature under the upper bound
\begin{equation}\label{eq:t_batt_bound}
\vartwo{T,batt} \leq \overline{\vartwo{T,batt}}
\end{equation}
to protect battery lifetime. For a change $\Delta\vartwo{T,batt}$ between intervals, battery temperature is predicted using the linear
\begin{equation}\label{eq:t_batt}
\varthree{T,batt,i+1}=\varthree{T,batt,i}+\Delta\varthree{T,batt,i}.
\end{equation}

Temperature changes are caused by electrochemical losses during use, heat emitted passively to the surroundings, and heat extracted by the active cooling system. The battery can be modeled as a lumped mass $\vartwo{m,batt}$ with thermal capacity $\vartwo{c,batt}$ that admits a thermal content change of $\vartwo{m,batt}\vartwo{c,batt}\Delta\vartwo{T,batt}$ for the change $\Delta\vartwo{T,batt}$ \cite{RN854}. Using the fictitious forces convention, the heat balance is
\begin{equation}\label{eq:heat_balance_1}
\vartwo{m,batt}\vartwo{c,batt}\Delta\vartwo{T,batt} = (\vartwo{Q,gen}-\vartwo{Q,rem})\vartwo{\Delta,s},
\end{equation} 
where $\vartwo{Q,gen}$ and $\vartwo{Q,rem}$ denote the battery heat generated and removed per meter traveled, respectively.

\subsubsection{Derivation of Heat Generated}

$\vartwo{Q,gen}$ can be expressed in terms battery efficiency for both charging and discharging
\begin{equation}\label{eq:f_gen_abs}
\vartwo{Q,gen} = |\vartwo{F,batt}|\big(1-\vartwo{\eta,batt}(\vartwo{F,batt},v)\big);
\end{equation} 
however, the linear equality \eqref{eq:heat_balance_1} would not remain linear if it were to admit the absolute value operation $|\vartwo{F,batt}|$. Alternatively, we propose to mimic $|\vartwo{F,batt}|$ using $\vartwo{F,dis}-\vartwo{F,chr}$ 
\begin{equation}\label{eq:f_gen_pos_neg}
\vartwo{Q,gen} = (\vartwo{F,dis}-\vartwo{F,chr})\big(1-\vartwo{\eta,batt}(\vartwo{F,batt},v)\big),
\end{equation} 
where $\vartwo{F,dis}\geq\vartwo{F,batt},0$ and $\vartwo{F,chr}\leq\vartwo{F,batt},0$. Section 3 explains how $\vartwo{F,dis}$ and $\vartwo{F,chr}$ adopt the positive discharging and negative charging values of $\vartwo{F,batt}$, respectively. Lastly, the variable efficiency term $\vartwo{\eta,batt}(\vartwo{F,batt},v)$ is simplified to the constant
\begin{equation}\label{eq:f_gen_pos_neg_eff}
\begin{split}
\vartwo{Q,gen} = &\vartwo{F,dis}(1-\vartwo{\widetilde{\eta},dis})-\vartwo{F,chr}(1-\vartwo{\widetilde{\eta},chr}),
\end{split} 
\end{equation}
where $\vartwo{\widetilde{\eta},dis}$ and $\vartwo{\widetilde{\eta},chr}$ denote average discharging and charging efficiency, respectively.

\subsubsection{Derivation of Heat Removed}

The heat removed from the battery per meter comprises of heat emitted to ambient per meter $\vartwo{Q,amb}$ and heat extracted by active cooling system per meter $\vartwo{Q,cool}$
\begin{equation}\label{eq:heat_removed}
\vartwo{Q,rem}=\vartwo{Q,amb}+\vartwo{Q,cool}.
\end{equation}

The heat lost to ambient per second is $h(\vartwo{T,batt}-\vartwo{T,amb})$, where $h$ is rate of heat transfer per second per kelvin. As such, $\vartwo{Q,amb}$ becomes
\begin{equation}\label{eq:heat_ambient}
\vartwo{Q,amb}= h(\vartwo{T,batt}-\vartwo{T,amb})\vartwo{\delta,t},
\end{equation}
where $\varthree{\delta,t,i}=1/v_i$ is the time required to travel one meter at the speed of interval $i$. In the space-domain, $\vartwo{\delta,t}$ is dependent on the optimized speed profile and thus unknown \textit{a priori}; therefore, the fixed average value $\vartwo{\widetilde{\delta},t}$ is used instead, except for station dwell (wait) times which are known \textit{a priori}. The error induced by this approximation is negligible.

A relation between $\vartwo{Q,cool}$ and $\vartwo{F,cool}$ is required for the balance expression \eqref{eq:force_balance_polynomial}. Assuming a direct connection between energy used and heat removed $\vartwo{F,cool} = \vartwo{Q,cool}/\widetilde{\gamma}$, where $\widetilde{\gamma}$ is the average coefficient of performance---the ratio of active cooling rate to power consumed by cooling system.

Lastly, peak heat removal rate per second $\overline{\vartwo{\dot{Q},cool}}$ is related to $\vartwo{Q,cool}$ by
\begin{equation}
\vartwo{Q,cool} \leq \overline{\vartwo{\dot{Q},cool}} \lambda_v.
\end{equation}

\subsubsection{Compilation of Thermal Model}

Substitute \eqref{eq:f_gen_pos_neg_eff} and \eqref{eq:heat_removed} into \eqref{eq:heat_balance_1} to get the linear
\begin{equation}\label{eq:compiled_thermal}
\begin{split}
\vartwo{m,batt}\vartwo{c,batt}\Delta\vartwo{T,batt}= \Big(&\vartwo{F,dis}(1-\vartwo{\widetilde{\eta},dis})-\vartwo{F,chr}(1-\vartwo{\widetilde{\eta},chr})\\
&-h(\vartwo{T,batt}-\vartwo{T,amb})\vartwo{{\widetilde{\delta}},t}-\vartwo{Q,cool} \Big) \vartwo{\Delta,s}.
\end{split}
\end{equation}

\section{Optimization Formulation}

The models derived in section 2 are now used to formulate the target optimization problem. The optimized system states are $(z,\zeta,\vartwo{T,batt})$; the main control variables are $(\vartwo{F,m},\vartwo{F,brk},\vartwo{F,fc},\vartwo{F,batt},\vartwo{Q,cool})$; and the auxiliary variables are $(v,\lambda_v,\lambda_\zeta,\Delta_\zeta,\Delta\vartwo{T,batt},\vartwo{F,dis},\vartwo{F,chr})$. After obtaining the optimal solution, the optimal trajectories of the fictitious variables $(\vartwo{F,fc},\vartwo{F,batt},\vartwo{Q,cool})$ are multiplied by speed in order to obtain their respective power commands, namely fuel cell power output, battery power output, and cooling rate per second.

The optimization problem computes the trajectory for $N$ intervals from $i=0,1,\cdots,N-1$ starting with initial states $(z_0,\zeta_0,\varthree{T,batt,0})$. The cost function penalizes hydrogen fuel
\begin{equation}\label{eq:cost_function}
\sum_i \vartwo{q,fc}(\varthree{F,fc,i},\vartwo{z,i})\varthree{\Delta,s,i}.
\end{equation}

The linear equality constraints \eqref{eq:kinetic_energy_2}, \eqref{eq:zeta}, and \eqref{eq:t_batt}, predict the system's states $(z,\zeta,\vartwo{T,batt})$, respectively. A second set of necessary equality constraints are \eqref{eq:lambda_zeta} and \eqref{eq:compiled_thermal} for the auxiliary variables $\lambda_\zeta$ and $\Delta\vartwo{T,batt}$. Moreover, the equality
\begin{equation}\label{eq:charge_sustain}
\zeta_N = \zeta_0
\end{equation}
enforces charge-sustaining operation on the battery,
\begin{equation}\label{eq:journey_time}
\sum_i \varthree{\Delta,s,i}\lambda_{v,i} = \tau
\end{equation}
terminates the journey exactly $\tau$ seconds after start, and
\begin{equation}\label{eq:station_stops}
v_j = \vartwo{v,stop}
\end{equation}
halts the train at station stop intervals denoted $j$.

The linear inequality constraints are broken down into the simple lower and upper bounds
\begin{subequations}
\begin{alignat}{2}
\varthree{F,chr,i} & \leq \hspace{1em} 0 && \leq \lambda_{v,i},\lambda_{\zeta,i}, \varthree{F,dis,i}\\
\underline{v} & \leq \hspace{0.9em} v_i && \leq \overline{v},\\ 
\underline{v}^2 & \leq \hspace{0.9em} z_i && \leq \overline{v}^2,\\ 
\underline{\zeta} & \leq \hspace{0.9em} \zeta_i && \leq \overline{\zeta},\\
& \hspace{1.5em} \varthree{T,batt,i} && \leq \overline{\vartwo{T,batt}},\label{eq:temp_bound_inequality}\\
\underline{\vartwo{F,m}} & \leq \hspace{0.2em} \varthree{F,m,i} && \leq \overline{\vartwo{F,m}},\\ 
\underline{\vartwo{F,brk}} & \leq \hspace{0.1em} \varthree{F,brk,i} && \leq \overline{\vartwo{F,brk}}
\end{alignat}
\end{subequations}
and the more elaborate linear inequalities
\begin{subequations}
\begin{alignat}{2}
\underline{\vartwo{P,m}}\lambda_{v,i} & \leq \varthree{F,m,i} && \leq \overline{\vartwo{P,m}}\lambda_{v,i},\\ 
\underline{\vartwo{P,batt}}\lambda_{v,i} & \leq \varthree{F,batt,i} && \leq \overline{\vartwo{P,batt}}\lambda_{v,i},\\ 
\underline{\vartwo{P,fc}}\lambda_{v,i} & \leq \varthree{F,fc,i} && \leq \overline{\vartwo{P,fc}}\lambda_{v,i},\\
0 & \leq \varthree{Q,cool,i} && \leq \overline{\vartwo{\dot{Q},cool}}\lambda_{v,i}.
\end{alignat}
\end{subequations}

Lastly are the list of relaxed inequalities
\begin{subequations}\label{eq:relaxed_constraints}
\begin{alignat}{2}
1 & \leq v_i \lambda_{v,i},\label{eq:relaxed_1}\\
v_i^2 & \leq z_i,\label{eq:relaxed_2}\\
\vartwo{q,m}(\varthree{F,m,i},z_i) + \varthree{Q,cool,i}/\widetilde{\gamma} + \vartwo{P,aux}\lambda_{v,i} & \leq \varthree{F,fc,i} + \varthree{F,batt,i},\label{eq:relaxed_3}\\
\alpha \varthree{F,batt,i}^2\varthree{\Delta,s,i} & \leq \lambda_{\zeta,i} \lambda_{v,i},\label{eq:relaxed_4}\\
\varthree{F,chr,i} & \leq \varthree{F,batt,i},\label{eq:relaxed_6}\\
\varthree{F,batt,i} & \leq \varthree{F,dis,i}.\label{eq:relaxed_7}
\end{alignat}
\end{subequations}

The constraint \eqref{eq:relaxed_1} implies that $v$ is strictly positive and thus $z$ as well due to \eqref{eq:relaxed_2}. Nevertheless, in order to emulate being stationary at station stops in \eqref{eq:station_stops}, $\vartwo{v,stop}$ is set to a small positive value that approaches zero. During station stops $\varthree{F,ext,j}$ is zeroed in order to successfully emulate a stationary state without motion induced resistance, see \eqref{eq:f_ext}. Since the optimized speed profile is strictly positive, the sampling intervals during station stops $\Delta_{\text{s},j}$ are adjusted \textit{a priori} to the multiplication of expected dwell time by $\vartwo{v,stop}$. Although the optimized speed at station stops never attains zero, in practice, it can be zeroed without affecting feasibility or optimality if $\vartwo{v,stop}$ in \eqref{eq:station_stops} was small enough.

In order to prove the optimality of the proposed formulation, the relaxed constraints \eqref{eq:relaxed_constraints} need to be proven to hold with equality. The following justifies inequality tightness:

\begin{itemize}
\item \eqref{eq:relaxed_1}: the summation $\sum_i \lambda_{v,i}$ is fixed through \eqref{eq:journey_time} and $v$ has the incentive to drop due to losses in \eqref{eq:f_ext};
\item \eqref{eq:relaxed_2}: $z$ has incentive to drop due to penalty \eqref{eq:cost_function} but $v$ is constrained by \eqref{eq:relaxed_1};
\item \eqref{eq:relaxed_3}: $\vartwo{F,batt}$ has incentive to drop or turn negative for free battery charge, $\vartwo{F,fc}$ is minimized by penalty \eqref{eq:cost_function}, whereas the left-hand side is necessary for auxiliary loads and to move the train for the journey time constraint \eqref{eq:journey_time};
\item \eqref{eq:relaxed_4}: the relaxed version of its original model \eqref{eq:F_batt_original}, $\Delta_\zeta \geq (\alpha\vartwo{F,batt}^2v^2 + \beta \vartwo{F,batt}v)\vartwo{\Delta,t}$, is tight when relaxed, as it would rather minimize $\Delta_\zeta$ or turn it negative to gain free battery charge, whereas $\vartwo{F,batt}$ on the right-hand side is necessary to move the train in \eqref{eq:relaxed_3};
\item \eqref{eq:relaxed_6},\eqref{eq:relaxed_7}: if the upper temperature bound \eqref{eq:temp_bound_inequality} is reached, \eqref{eq:compiled_thermal} would rather tighten \eqref{eq:relaxed_6} and \eqref{eq:relaxed_7} before relying on the active cooling system command $\vartwo{Q,cool}$ that is indirectly penalized by \eqref{eq:cost_function} through \eqref{eq:relaxed_3}. Tightness is only guaranteed if the upper temperature bound is reached otherwise the thermal constraint is not necessary. Tightness is not guaranteed for a lower bound on temperature but that is not likely needed for traction batteries during use.
\end{itemize}

The optimization formulation proposed is convex because it penalizes a convex quadratic cost function subject to linear equality and convex inequality constraints. We formulate it as a second-order cone program and solve it using the barrier method \cite{gurobi}.

\section{Simulation Results}

The formulation proposed in section 3 is now tested on a benchmark train and rail journey under three different ambient temperatures: \SI{-5}{\celsius}, \SI{20}{\celsius} and \SI{35}{\celsius}. The target train optimized is the \textit{HydroFLEX} \cite{hydroflex}, a four-car hydrogen train with parameters shown in Table \ref{tab:train_parameters}. The rail line simulated is the $63$-km-long \textit{Tees Valley Line} located in northern \textit{England}. The line runs between \textit{Saltburn} and \textit{Bishop Auckland} with 16 intermediate stops. Route elevation data has been extracted from \cite{elevation} using the EU-DEM dataset. The route is optimized and simulated at a spatial sampling interval $\Delta_\text{s}$ of \SI{10}{\meter} leading to around 6300 intervals. 

In order to guarantee a fair comparison, the following has been unified between simulations: initial and terminal state-of-charge of $50\%$ using \eqref{eq:charge_sustain}; total journey time of 87 minutes using \eqref{eq:journey_time}; dwell time at stations. The initial battery temperature is assumed equal to ambient for \SI{20}{\celsius} and \SI{35}{\celsius}, whereas the battery is pre-heated to \SI{10}{\celsius} for the ambient \SI{-5}{\celsius} in order to prepare it for peak performance. The upper temperature bound $\overline{\vartwo{T,batt}}$ is \SI{40}{\celsius}.

\begin{table}
\centering
\setlength{\tabcolsep}{3.5pt} %horizontal spacing
\renewcommand{\arraystretch}{1.7} %vertical spacing
\caption{Simulated Train Parameters}
\label{tab:train_parameters}
\resizebox{\columnwidth}{!}{
\begin{tabular}{rl rl rl rl}
\hline
\multicolumn{2}{l}{\hspace{2em}Vehicle}&\multicolumn{2}{l}{\hspace{2em}Motor}&\multicolumn{2}{l}{\hspace{2em}Battery}&\multicolumn{2}{l}{\hspace{2em}Cooling}\\
\hline
$m$&$\SI{183}{\tonne}$&$\underline{\vartwo{P,m}}$&$\SI{-585}{\kW}$&$\underline{\vartwo{P,batt}}$&$\SI{-600}{\kW}$&$\overline{\vartwo{T,batt}}$&\SI{40}{\celsius}\\
$\lambda$&$0.0625$&$\overline{\vartwo{P,m}}$&$\SI{585}{\kW}$&$\overline{\vartwo{P,batt}}$&$\SI{600}{\kW}$&$\overline{\vartwo{\dot{Q},cool}}$&\SI{15}{\kilo\watt}\\
$a$&$\SI{1,743}{\kN}$&$\underline{\vartwo{F,m}}$&$\SI{-87}{\kN}$&$Q$&$\SI{220}{\kWh}$&$\widetilde{\gamma}$&4\\
$b$&$\SI{76.4}{\kilogram\per\second}$&$\overline{\vartwo{F,m}}$&$\SI{87}{\kN}$&$R$&$\SI{21.7}{\mohm}$&$\vartwo{m,batt}$&\SI{3}{\tonne}\\
$c$&$\SI{6.2}{\kilogram\per\meter}$&$\underline{\vartwo{P,fc}}$&$\SI{24}{\kW}$&$\vartwo{U,oc}$&$\SI{600}{\volt}$&$\vartwo{c,batt}$&\SI{1}{\kilo\joule\per\kg\per\kelvin}\\
$\underline{\vartwo{F,brk}}$&$\SI{-180}{kN}$&$\overline{\vartwo{P,fc}}$&$\SI{400}{\kW}$&$\underline{\zeta}$&$20\%$&$\vartwo{\eta,dis}$&0.9\\
$\vartwo{P,aux}$&$\SI{85}{\kW}$& & &$\overline{\zeta}$&$80\%$&$\vartwo{\eta,chr}$&0.9\\
 & & & & & &$\vartwo{h,amb}$&\SI{25}{\joule\per\second\per\kelvin}\\
\hline
\end{tabular}
}
\end{table}

\begin{figure}
\centering
\includegraphics[width=\columnwidth]{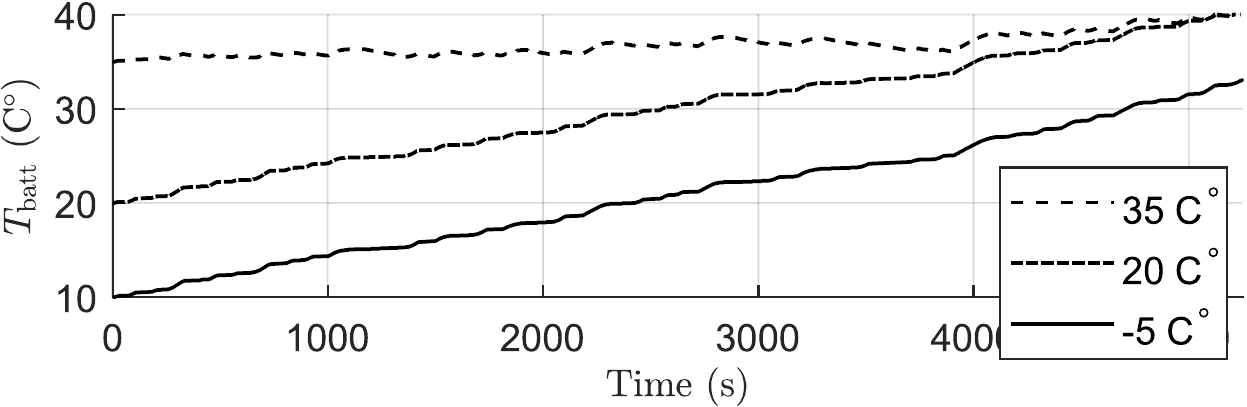}
\caption{Progression of battery temperature under three ambient conditions.}
\label{fig:all_temps}
\end{figure}

Figure \ref{fig:all_temps} shows the optimized battery temperature under all ambient conditions. The coldest profile \SI{-5}{\celsius} was the only not to reach the upper bound and not to require the active cooling system thus its trajectory is identical to previous formulations that do not optimize temperature \cite{mine_vppc}. Both warmer profiles required active cooling to remain under the upper temperature bound which led to higher fuel consumption, $0.7\%$ for \SI{20}{\celsius} and $1.7\%$ for \SI{35}{\celsius}.

\begin{figure}
\centering
\includegraphics[width=\columnwidth]{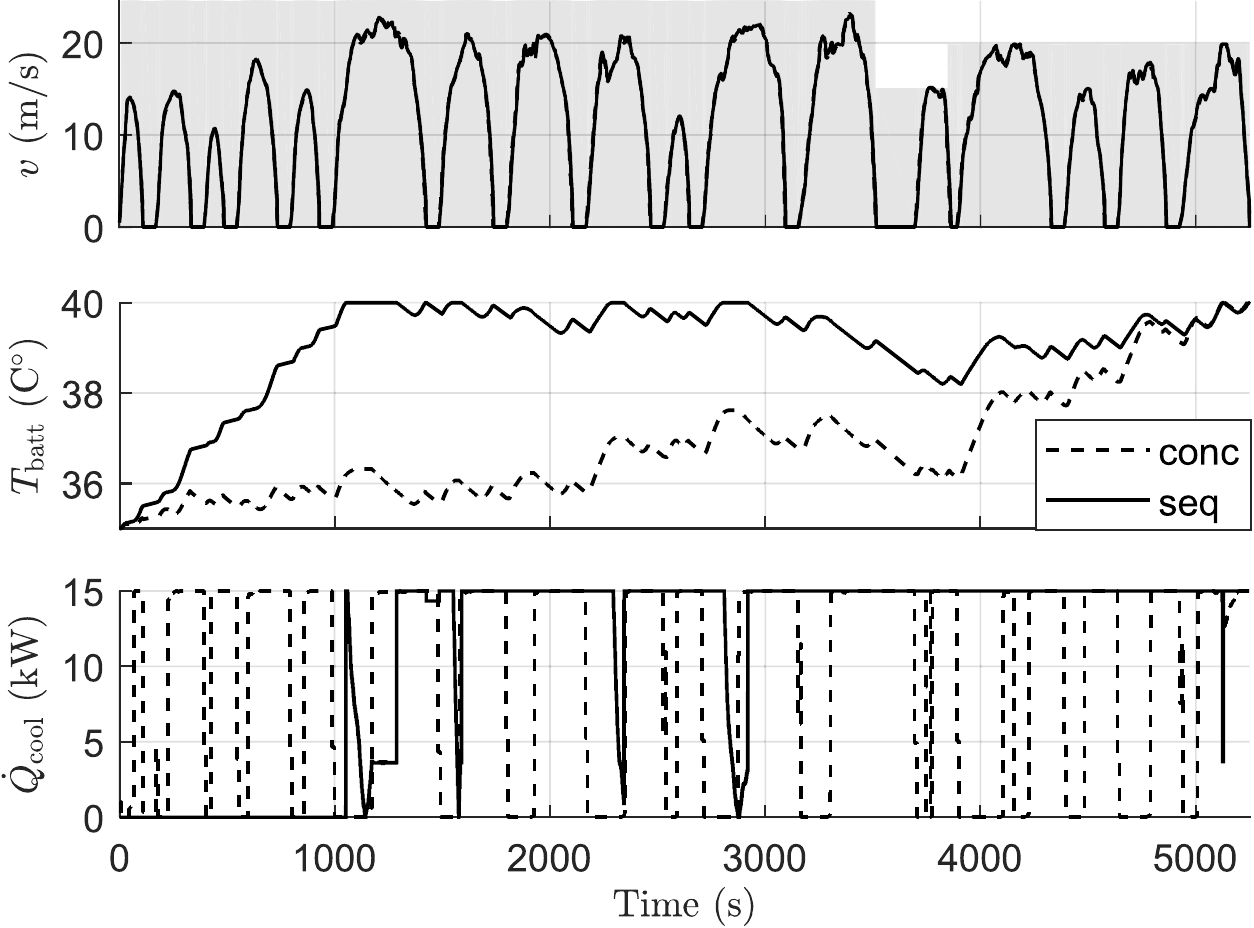}
\caption{Impact of method on cooling load in hot ambient of \SI{35}{\celsius}}
\label{fig:temp_seq_conc}
\end{figure}

Figure \ref{fig:temp_seq_conc} compares at an ambient of \SI{35}{\celsius} the concurrent method---all trajectories are optimized together as in section 3---against the sequential method---only speed and EMS are optimized together, whereas temperature is optimized later during operation. The concurrent is able to control battery temperature more effectively with less cooling effort and thus cost $3\%$ less fuel. This thermal enhancement is partly due to the concurrent using cooling early-on before temperature became too hot and partly because it generated less heat to begin with. The concurrent's fuel benefit to the sequential at \SI{20}{\celsius} ambient was a more modest $0.3\%$ due to the lower active cooling demand in a cooler environment.

\begin{figure}
\centering
\includegraphics[width=\columnwidth]{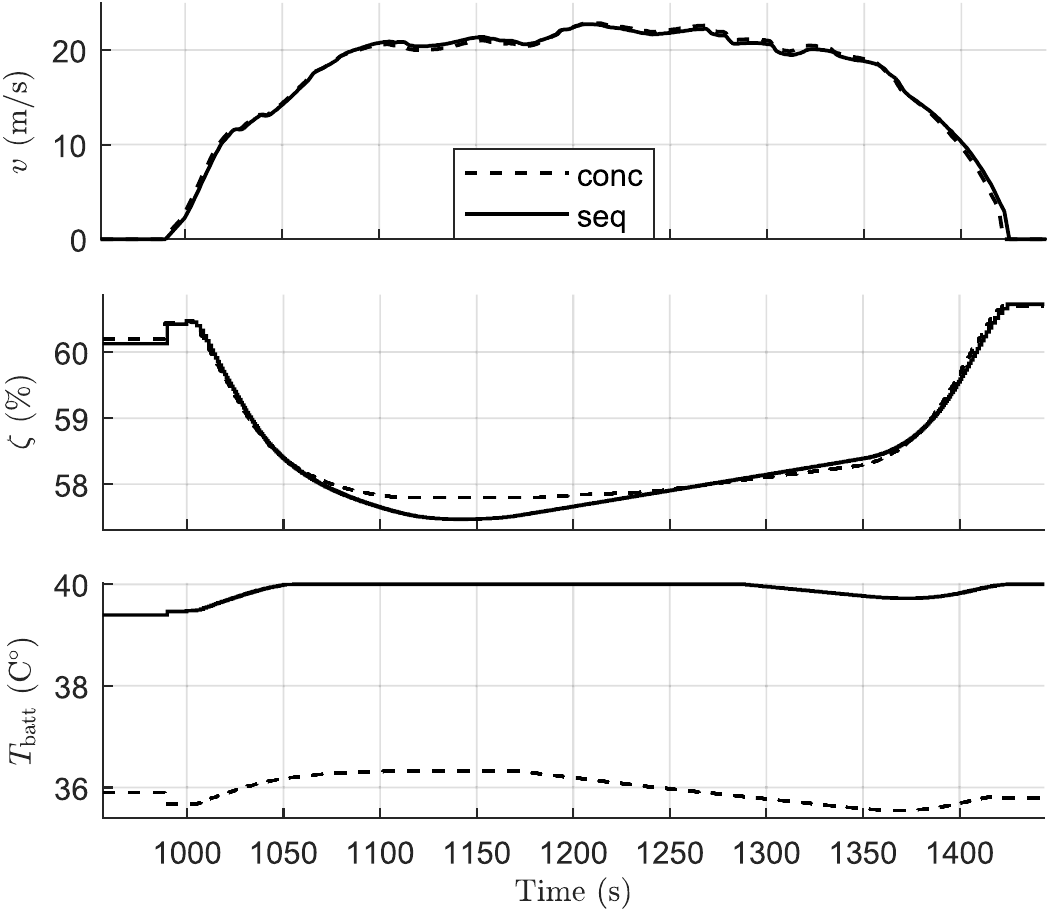}
\caption{Impact of method on battery utilization in hot ambient of \SI{35}{\celsius}.}
\label{fig:power_seq_con}
\end{figure}

Figure \ref{fig:power_seq_con} gives an example of why the concurrent generated less heat than the sequential using trajectories between two stations. The state-of-charge plot $\zeta$ shows the concurrent's battery being discharged less than the sequential at around \SI{1100}{\second} which subsequently required less recharging later on. This lower utilization of the concurrent's battery leads to less temperature rise and is repeated throughout. While this implies that the fuel cell compensates for the battery's lower utilization, the concurrent finds a global minima that balances between battery utilization and battery cooling consumption.

\section{Conclusion}

A convex formulation for concurrently optimizing train speed, EMS, and battery temperature is proposed. It is concluded that optimizing battery temperature concurrently with speed and EMS can enhance both fuel consumption and thermal management. The degree of benefit is reliant on ambient temperature. Quick convex optimization can empower the proposed formulation to be used in real-time in order to get more accurate ambient readings. A continuation to this work is to investigate fuel cell temperature.

\bibliographystyle{IEEEtran}
\bibliography{ecc2022_bib}

\end{document}